# Hybrid Voltage and Current Control Method for Harmonic Mitigation of Single-Phase AC Loads in DC Microgrids

Mehdi Baharizadeh, *Senior Member*, *IEEE*, Mohammad Sadegh Golsorkhi, *Senior Member*, *IEEE*, Neda Keshavarzi, and Thomas Ebel, *Senior Member*, *IEEE*

*Abstract*— DC microgrids provide an efficient framework for interconnection of DC distributed energy resources (DERs) and DC loads. To continue to supply legacy single phase AC loads, DC/AC converters can be integrated in the DC microgrid. The oscillatory instantaneous power of the single-phase AC load translates into a harmonic current on the converter's DC side, which increases the losses and causes unwanted voltage harmonics in the DC microgrid. To mitigate this issue, this paper proposes a hybrid voltage and current control method (HCM) for DERs. This scheme consists of an inner current control loop and an outer control layer which determines the reference for the inner loop. The outer control layer combines the DC voltage control loop with an output harmonic current control loop. This hybrid structure enables simultaneous regulation of the DC components of the DER output voltage and control of the harmonic component of the DER output current in accordance with the local single-phase AC load's demand. Frequency-domain analysis of the proposed method is presented to demonstrate the DC voltage and harmonic current loops are decoupled and there is no unwanted interaction between them. Additionally, time-domain response of the proposed scheme is validated through hardware-in-the-loop test results.

*Index Terms*—Current control, DC microgrids, harmonic compensation, voltage control.

## I. INTRODUCTION

Microgrids are parts of electrical distribution systems that include distributed energy resources (DERs) and loads capable of operating autonomously as an island. Since traditional distribution systems are AC, AC microgrids have been widely adopted [1]. However, the fact that the energy resources in DERs are commonly of DC nature (e.g., PV, battery) and the utilization of DC loads (e.g., LED lights, electronic loads, electric vehicles) is becoming increasingly common, has resulted in surging interest in DC distribution systems and DC microgrids [2, 3]. The DC microgrid configuration enables eliminating extra DC/AC and AC/DC converter stages in DC DERs and DC loads, respectively, hence offering higher efficiency, improved reliability, and reduced costs [4, 5].

DC microgrids operate in two modes: grid-connected and islanded. In grid-connected mode, the converter linking the

M. Baharizadeh (*corresponding author*), Mohammad Sadegh Golsorkhi, and Thomas Ebel are with the Centre for Industrial Electronics, Institute of Mechanical and Electrical Engineering, University of Southern Denmark, Sonderborg, Denmark (corresponding author's email: smbaharizadeh@sdu.dk). Neda Keshavarzi is with The Department of Electrical Engineering Khomeinishahr Branch, Islamic Azad University Isfahan, Iran.

microgrid to the AC grid regulates the DC microgrid voltage through proper power exchange between the DC microgrid and the upstream AC grid. In this mode, depending on the DER technology, different strategies such as maximum power point tracking and storage charge management are applied, all of which utilize the current-controlled method (CCM) for the DER's grid-side converter [6]. By employing CCM, DERs can also compensate their local loads, similar to the performance of active filters [7]. Islanded mode is activated by disconnecting from the upstream AC grid. In this mode, which is the focus of this paper, DERs are responsible for regulating the DC microgrid voltage [8, 9]. Accordingly, DERs' grid-side converter operate using a voltage-controlled method (VCM). In this mode, DERs must balance generation and consumption, sharing the DC microgrid load among themselves according to their capacities. To achieve this power management strategy without relying on a communication network, droop characteristics are commonly utilized [6, 10].

Despite the trend towards increasing penetration of DC loads, the AC loads must still be considered in DC microgrids [11, 12]. In the case of low-power DC microgrids, such as residential microgrids, where AC loads are single-phase, a single-phase inverter can be used to supply the AC loads [13]. The instantaneous power of single-phase AC loads oscillates at twice the fundamental frequency, producing a harmonic current of the same frequency on the DC side of the inverter [14], referred to as the second harmonic current throughout this paper. Although the second harmonic current is filtered and partly absorbed by the inverter's DC link capacitor, the study in [15] shows that a substantial second harmonic current can still be injected into the DC lines. Two main approaches can be considered to address this harmonic current. The first, discussed in [15, 16], seeks to share the second harmonic current among DERs to avoid overstressing any single unit. However, this results in the harmonic current flowing through the distribution lines, increasing losses and second harmonic voltage ripple. The second approach, which is the focus of this paper, compensates the harmonic current locally at DER's single-phase load. This significantly reduces harmonic current flow through the lines, thereby lowering losses and improving DC voltage quality. Moreover, local second harmonic compensation enables reducing the size of the DC-link capacitor in the single-phase inverter.

To achieve these objectives, DER's grid-side converter in DC microgrid must perform two tasks simultaneously: (1) operate in VCM to regulate the DC component of their terminal voltage, which is essential for islanded operation, and (2) employ CCM to regulate the harmonic component of their output current, which is required to compensate the second harmonic current associated with the local single-phase load. Realizing VCM and CCM at different frequencies, known as hybrid voltage and current control method (HCM), has already been studied for AC DERs. In [7], an HCM is proposed for DER's inverter in an islanded AC microgrid that regulates the fundamental component of the terminal voltage along with harmonic current output for local nonlinear load compensation. However, this method eliminates the inner current control loop. The existence of an inner current control loop enables the control system to apply an upper limit on the current throughout short-circuit faults and grid disturbances. This feature is particularly useful for sustained operation of the converter during such disturbances, hence availing implementation of low voltage ride through (LVRT) [17, 18]. In [19], a converter connected to an islanded AC microgrid uses HCM to regulate the fundamental component of the output current, as well as the harmonic component of the terminal voltage. In this approach, the harmonic voltage reference is set to zero to eliminate harmonic voltage distortion at the converter terminal, resulting in negligible harmonic impedance and automatic compensation of local nonlinear load current. However, this method lacks control over the amount of current allocated for harmonic compensation and requires another source to regulate the fundamental component of the AC microgrid voltage. An HCM with the same control objectives in [19] is applied to a grid-connected inverter in [20]. However, this method also removes the inner current control loop, making the inverter's semiconductor switches vulnerable to overcurrent during grid faults.

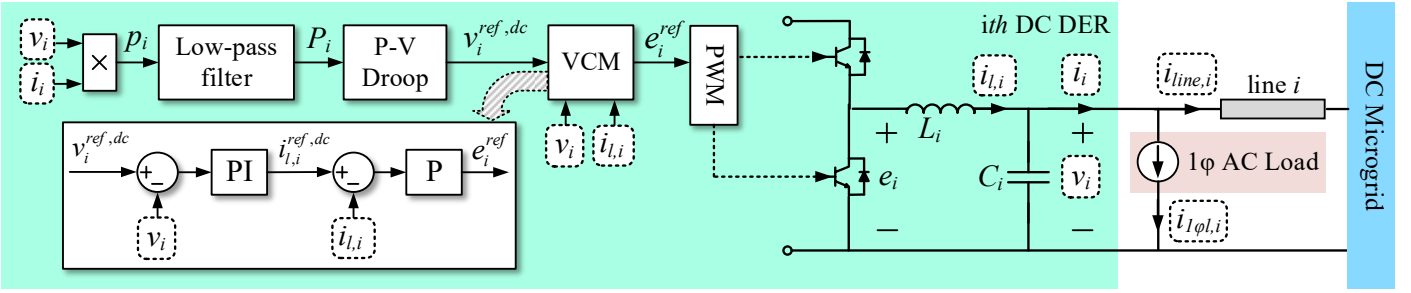

Fig. 1. DC DER grid-side converter and its control schematic, including VCM for regulating the DC component of output voltage

Despite existing studies for AC DERs, HCM for DC DERs has not been studied in detail by other researchers. This paper proposes an HCM for DC DERs that simultaneously regulates the DC component of the terminal voltage and the harmonic component of the output current. The method combines two decoupled outer loops, proportional-Integral (PI) controller-based DC voltage regulation, and Resonance (R) controller-based harmonic current regulation, along with an inner current loop using a Proportional (P) controller for filter inductor current control. The main contributions of this paper are as follows:

- In contrast to the methods in [15, 16], which share the second harmonic current among DC DERs, the proposed method compensates this harmonic current locally, preventing its flow through the lines, reducing losses, and improving DC voltage quality.
- Compared to HCM approaches in [5, 19, 20] for AC DERs, this work focuses on HCM for DERs in DC systems and addresses their specific control design requirements.
- Unlike the HCM methods in [7, 20], the proposed HCM control strategy has an inner current control loop. This inner current control loop facilitates applying a current limit to the DERs during short circuit faults within the microgrid, hence enabling sustained operation amid temporary faults.
- Compared to the conference paper [21], of which this paper is an extension, this paper provides comprehensive frequency-domain simulations and hardware-in-the-loop (HIL) time-domain experiments.

The paper is organized as follows: To provide a clear understanding of the proposed HCM, the conventional VCM and CCM are first described. Section II presents the VCM used to regulate the DC component of the DER's terminal voltage. Section III explains the CCM for regulating the harmonic component of a converter's output current in a DC grid. Section IV details the proposed HCM as a combination of these two methods. Frequency-domain simulations and HIL-based time-domain experiments are presented in Section V, and conclusions are drawn in Section VI.

## II. VCM for Voltage DC Component Regulation

In islanded mode, DERs are responsible for regulating the DC component of voltage, supplying the loads, and sharing the load power among themselves. Droop characteristics are commonly used to achieve these objectives based on a decentralized control method. The P-V droop characteristic of DERs is as follows:

$$v_i^{ref,dc} = V_{max} - m_i . P_i \tag{1}$$

$$m_i = \frac{V_{max} - V_{min}}{P_{i,max}} \tag{2}$$

where $v_i^{ref,dc}$ is the DC component reference value of the DER's terminal voltage, $m_i$ is the droop slope, $P_i$ is the lowpass-filtered power generation of the $i$th DER, and $V_{max}$, $V_{min}$ are the maximum and minimum permissible voltage limits for the DC microgrid [4, 10].

The DC DER grid-side converter and its control schematic are shown in Fig. 1. As illustrated, the DER's instantaneous power $p_i$ is calculated from the measured output voltage and current and, after low-pass filtering, is applied to the P–V droop characteristic.

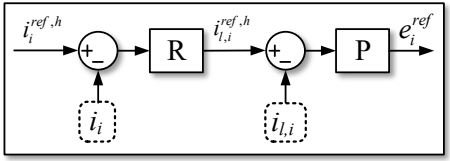

Fig. 2. Control schematic of CCM for regulating the harmonic component of output current

The droop equation calculates the DC component reference value of the DER's terminal voltage ($v_i^{ref,dc}$), which will be realized by employing VCM. The VCM consists of an outer voltage loop and an inner current loop. In the outer voltage loop, the difference between $v_i^{ref,dc}$ and the measured terminal voltage ($v_i$) is applied to a PI controller, which reacts to the DC component of error and calculates the reference value for the DC component of the filter inductor current ($i_{l,i}^{ref,dc}$). In the inner current loop, the difference between $i_{l,i}^{ref,dc}$ and the measured filter inductor current ($i_{l,i}$) is applied to a P controller, which generates the reference converter-side voltage ($e_i^{ref}$) required to regulate the DC component of the terminal voltage. The inner loop, based on a P controller, operates at a high bandwidth, improving the dynamic performance of the VCM.

## III. CCM FOR CURRENT HARMONIC COMPONENT REGULATION

The instantaneous power of single-phase AC loads oscillates at twice the fundamental frequency. When supplied via a single-phase inverter in DC microgrids, this oscillating power produces a second harmonic current on the DC side [15]. The presence of this harmonic current in the power lines increases losses and introduces a second harmonic voltage ripple. Such a harmonic component can be compensated by a grid-connected converter located near the load. To achieve this, it is sufficient for the converter's output current ($i_i$) to match the harmonic component of the single-phase AC load current on the DC side ($i_{1\varphi l,i}^h$). As a result, the line current ($i_{line,i}$) will not contain any second harmonic component. For this reason, the converter must employ the CCM, whose control schematic for the converter topology shown in Fig. 1 is illustrated in Fig. 2. As seen, the control schematic includes outer and inner current control loops. In the outer current loop, the difference between the reference harmonic component of the converter's output current ($i_i^{ref,h}$) and the measured output current ($i_i$) is applied to an R controller, whose transfer function is as follows:

$$R(s) = \frac{2k_i . \omega_c . s}{s^2 + 2\omega_c . s + {\omega_0}^2} \tag{3}$$

where $k_i$ is the resonant gain, $\omega_0$ is the frequency of the intended harmonic component (twice the fundamental frequency of the single-phase AC load), and $\omega_c$ is the bandwidth of the R controller. With its high gain at the intended frequency, the R controller reacts to the error at that frequency and determines the reference value for the harmonic component of the filter inductor current ($i_{l,i}^{ref,h}$). In the inner loop, the difference between $i_{l,i}^{ref,h}$ and $i_{l,i}$ is applied to a P controller, which generates the reference converter-side voltage ($e_i^{ref}$) required to regulate the harmonic component of the converter's output current. It is noteworthy that the harmonic compensation method presented in this section does not require the extraction of $i_{1\varphi l,i}^h$. Since the R controller responds only to the intended harmonic component, $i_i^{ref,h}$ can simply be set equal to the entire single-phase AC load current on the DC side ($i_{1\varphi l,i}$). Therefore, the presented method does not impose a high computational burden.

## IV. HYBRID CONTROL METHOD FOR SIMULTANEOUS REGULATION OF VOLTAGE DC AND CURRENT HARMONIC COMPONENTS

The control objective in this section is the simultaneous regulation of the DC component of the terminal voltage and the harmonic component of the output current. Achieving this allows DERs in an islanded DC microgrid to compensate the second harmonic current of their local single-phase AC load. The schematic of the proposed control is illustrated in Fig. 3. As illustrated, the structure includes two outer loops for voltage and current control and an inner current loop. Comparison of Figs. 1

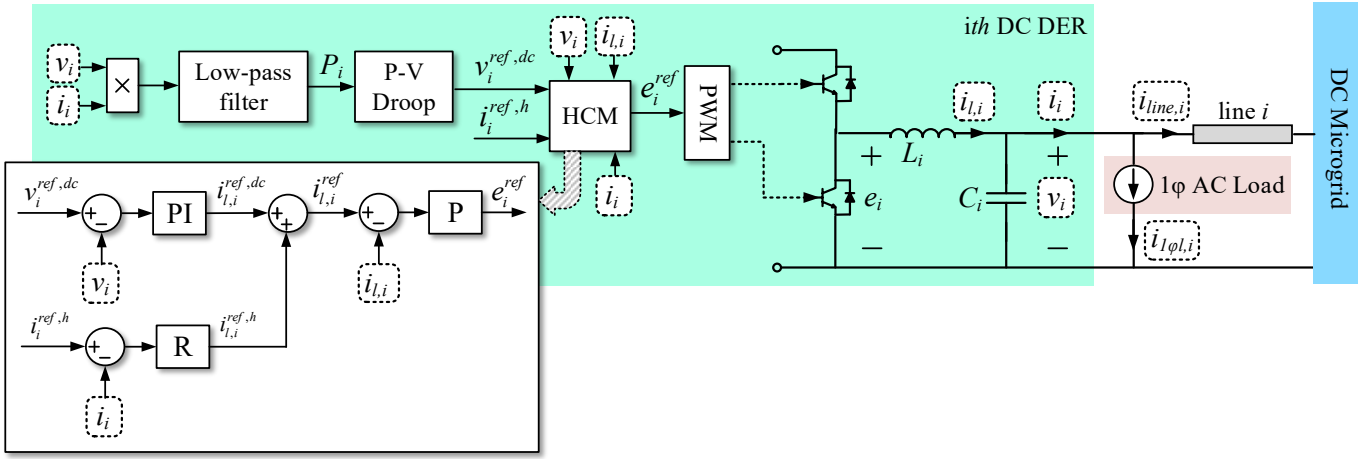

Fig. 3. Schematic diagram of DC DER grid-side converter with HCM control for simultaneous regulation of voltage DC component and current harmonic component

and 2 with Fig. 3 shows that the proposed control method combines the two previous approaches presented in sections II and III. The outer voltage loop in the proposed method, similar to the VCM described in Section II, employs a PI controller to generate the reference value for the DC component of the filter inductor current ($i_{l,i}^{ref,dc}$), which is required to regulate the DC component of the DER's terminal voltage. The inner current loop, similar to the CCM approach in Section III, uses an R controller to calculate the reference value for the harmonic component of the filter inductor current ($i_{l,i}^{ref,h}$) which is required to regulate the harmonic component of the output current. The reference values from the outer voltage and current loops are then combined to obtain the total reference current of the filter inductor ($i_{l,i}^{ref}$). Relying on Superposition, realizing this reference current enables simultaneous regulation of both the DC component of the terminal voltage and the harmonic component of the output current. $i_{l,i}^{ref}$ is subsequently applied to the inner current control loop, which determines the reference converter-side voltage ($e_i^{ref}$) required to achieve both control objectives. It is noteworthy that the outer voltage and current loops do not interfere with one another. The PI controller in the voltage loop has negligible gain at the intended harmonic frequency due to its small proportional gain, while the R controller in the current loop has negligible gain at DC. Additionally, similar to the CCM method described in Section III, in the proposed HCM the reference harmonic current ($i_i^{ref,h}$) can simply be set equal to the entire single-phase AC load current on the DC side ($i_{1\varphi l,i}$), as the R controller responds only to the intended harmonic component. Consequently, the extraction of the harmonic component of the load current is not required, and the computational burden remains low.

## V. Frequency-domain and HIL-based Time-domain Studies

In this section, frequency-domain analysis and HIL-based time-domain experimental studies of a test DC microgrid utilizing the proposed control method are presented to validate the conducted studies. The test DC microgrid is shown in Fig. 4. It includes two DERs, each connected to the point of common coupling (PCC) through a power line. A DC constant power load (CPL) is connected to the PCC, and a single-phase 50 Hz AC load, supplied through an inverter, is connected to the terminal of DER 1. No capacitor is used in the inverter's DC link in order to maximize the second harmonic current amplitude on the DC side, thereby evaluating the effectiveness of the proposed control under the worst physical condition. Both DERs implement droop characteristics. DER 1 employs HCM to realize the DC voltage reference determined by the droop characteristic and to compensate the second harmonic current (100 Hz component) associated with the single-phase AC load, while DER 2 employs VCM. The electrical and control parameters of the test DC microgrid are given in Tables I and II, respectively.

### A. Frequency-domain Analysis

In this subsection, the frequency-domain analysis of the voltage and current control for the proposed HCM is presented. The analysis is based on the frequency response of the closed-loop transfer functions of the HCM applied to DER 1 of the test system,

TABLE I. Electrical Parameters of the Test DC Microgrid

| Description | Value |
|---|---|
| DC DER 1 & 2 | $P_{max}$=10 kW, $V_{max}$=630 V, $V_{min}$=570 V |
| Line 1 & 2 | $R$=0.4 Ω, $L$=0.4 mH |
| DC CPL | 0s→5s: 6kW, 5s→10s: 12kW |
| Single-Ph. AC Load | 1.8 kW, $PF$=0.5, $f$=50 Hz |

TABLE II. Control Parameters of the Test DC Microgrid

| Description | Value |
|---|---|
| DC DER 1 | PI: $k_p$=0.4, $k_i$=50<br>R: $k_r$=30, $\omega_c$=5, $\omega_0$=628.32<br>P: $k_p$=5 |
| DC DER 2 | PI: $k_p$=0.4, $k_i$=50<br>P: $k_p$=5 |
| Cut-off Frequency of Low-pass Filters: 31.4 rad/s | |

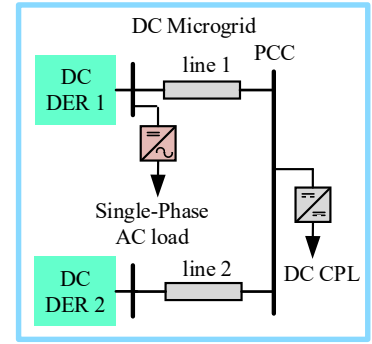

Fig. 4. Test DC microgrid

obtained using the AC sweep method in PSIM software. In the s-domain, the output current of DER1's ($I_1(s)$) can be expressed in terms of HCM inputs, which are reference values for terminal voltage and output current ($I_1^{ref,h}(s)$ and $V_1^{ref,dc}(s)$), as follows:

$$I_1(s) = G_{ii}(s) * I_1^{ref,h}(s) + G_{vi}(s) * V_1^{ref,dc}(s) + I_{dist}(s) \quad (4)$$

where $I_{dist}(s)$ represents variations in the output current caused by external disturbances such as load changes. $G_{ii}(s)$ and $G_{vi}(s)$ are closed-loop transfer functions, from $I_1^{ref,h}(s)$ and $V_1^{ref,dc}(s)$ to $I_1(s)$, respectively. The Bode plot of $G_{ii}(s)$ in Fig. 5 shows a 0 dB magnitude and zero phase at 100 Hz, indicating that under the proposed HCM, the output current accurately tracks its reference at the second harmonic frequency. As expected, the magnitude decreases at frequencies above and below 100 Hz, consistent with the R controller's low gain away from the target harmonic component. Figure 6 shows the Bode plot of $G_{vi}(s)$. At lower frequencies, $G_{vi}(s)$ shows a low-pass filtering characteristic, allowing the reference value for the terminal voltage impact the output current at DC but blocking it from causing high-frequency current ripples. At 100Hz, the magnitude shows a strong attenuation (around -30dB), indicating that the reference value of terminal voltage ($V_1^{ref,dc}$), which acts as a disturbance for current control, has negligible influence on 100 Hz harmonic current

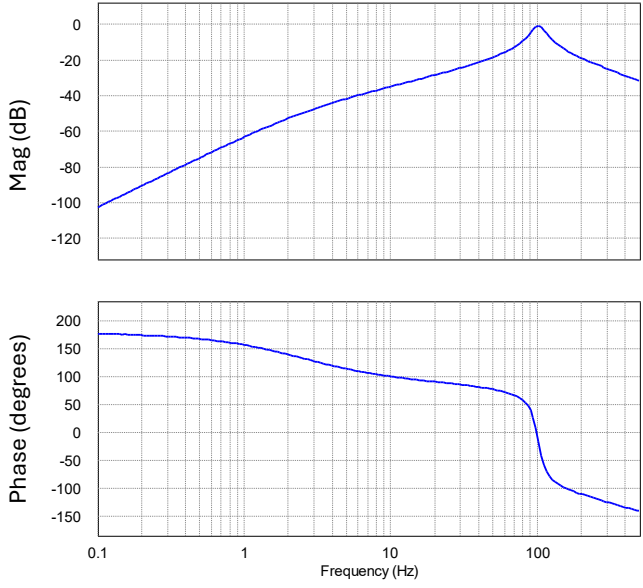

Fig. 5. Bode plot of $G_{ii}(s)$

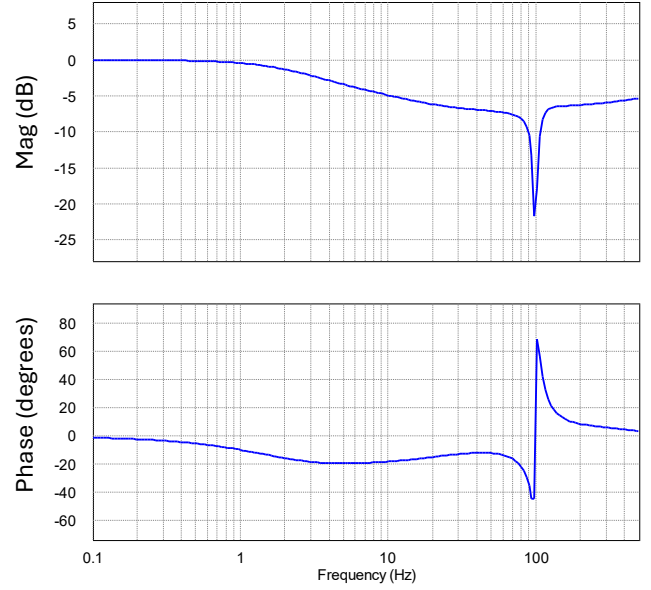

Fig. 7. Bode plot of $G_{vv}(s)$

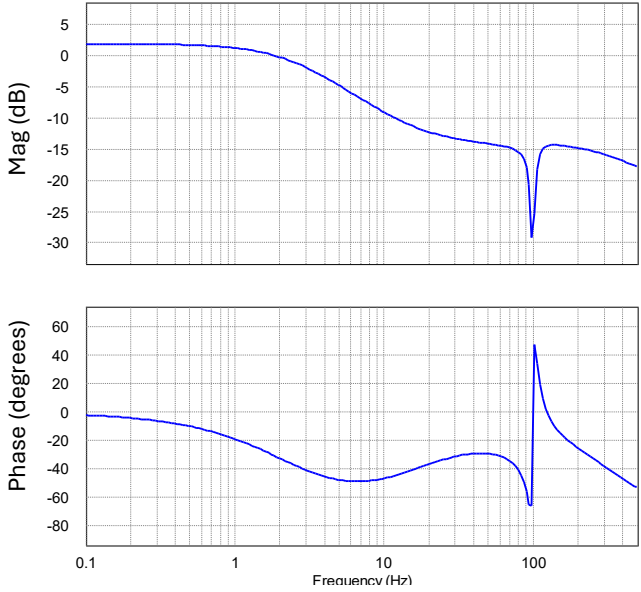

Fig. 6. Bode plot of $G_{vi}(s)$

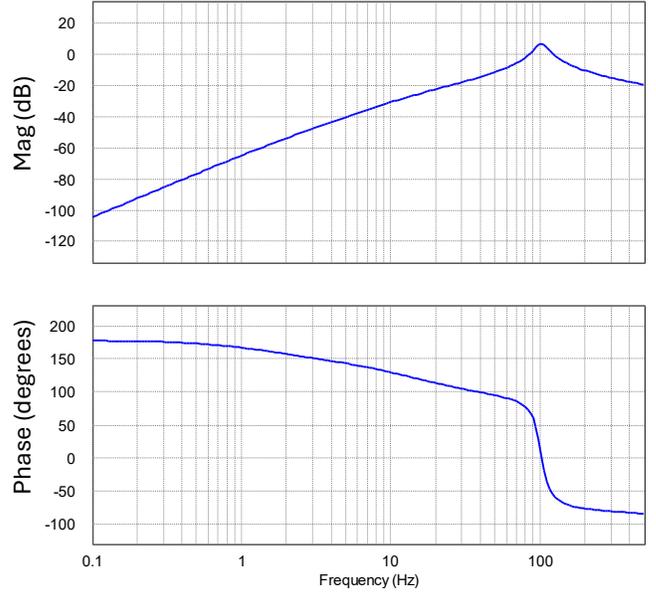

Fig. 8. Bode plot of $G_{iv}(s)$

tracking.

The terminal voltage of DER 1 ($V_1(s)$) can be expressed in terms of HCM inputs:

$$V_1(s) = G_{vv}(s) * V_1^{ref,dc}(s) + G_{iv}(s) * I_1^{ref,h}(s) \\ + V_{dist}(s) \qquad (5)$$

where $V_{dist}(s)$ is the changes in the terminal voltage caused by external disturbances such as load variations. $G_{vv}(s)$ and $G_{iv}(s)$ are the closed-loop transfer functions from $V_1^{ref,dc}(s)$ and $I_1^{ref,h}(s)$ to $V_1(s)$, respectively. The Bode plot of $G_{vv}(s)$ in Fig. 7 shows that at low frequencies, the magnitude is 0 dB with zero phase, confirming that the terminal voltage tracks its DC reference effectively. Strong attenuation at 100 Hz (-22dB) indicates negligible voltage-loop action at that frequency. Figure 8 shows the Bode plot of $G_{iv}(s)$. At low frequencies, the magnitude is negligible (around -100 dB), indicating that the reference value for output current ($I_1^{ref,h}$), which acts as a disturbance to voltage regulation, has negligible effect on DC voltage tracking. At 100Hz, the magnitude peaks at around 5dB, letting the reference value for the output current impact the terminal voltage at that component, which is related to the impedance seen by DER1. The minimal influence of the voltage reference on output-current tracking at 100 Hz, together with the negligible influence of the current reference on DC voltage tracking, demonstrates that the interaction between the voltage and current control loops is minimal.

### B. HIL-Based time-domain experimental results

In this sub-section, HIL experimental results obtained using an OPAL OP5700 lab setup (shown in Fig. 9) are presented. The current at the input of the DC/AC converter supplying the single-phase AC load is comprised of a 3A DC and a 6A harmonic component (100 Hz). Initially, the DC CPL draws 6 kW from the microgrid. To demonstrate the impact of load change on the system dynamics, the CPL's power is stepped up to 12 kW at t=5s. The DER1 is programmed such that initially, its outer harmonic current control loop is inactive (VCM control mode). At t=3s, DER1 switches to HCM control mode by activating its outer harmonic current loop. From t=3s to 7s, the HCM tries to fully compensate the second harmonic current of the single-phase AC load. Thereafter, from t=7s to 10s, DER 1 is programmed to compensate 50% of the harmonic current. Such partial compensation of the harmonic current may practically occur to prevent overcurrent in the DER converter when its current approaches the rated value.

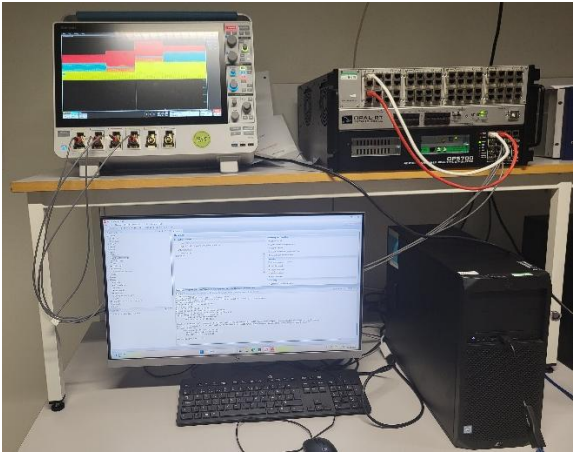

Fig. 9. OPAL-RT lab setup

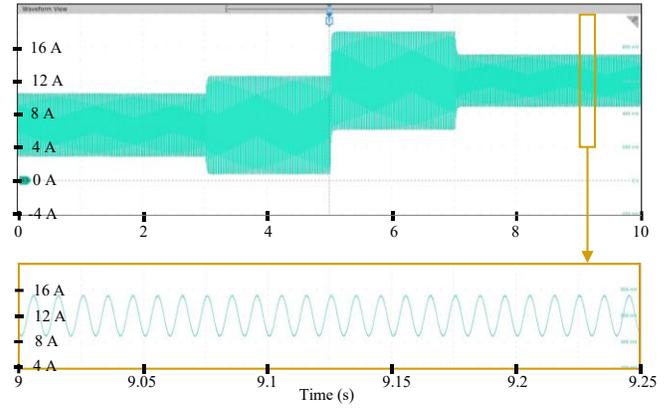

Fig. 11. HIL experimental results: output current of DER 1 ($i_1$).

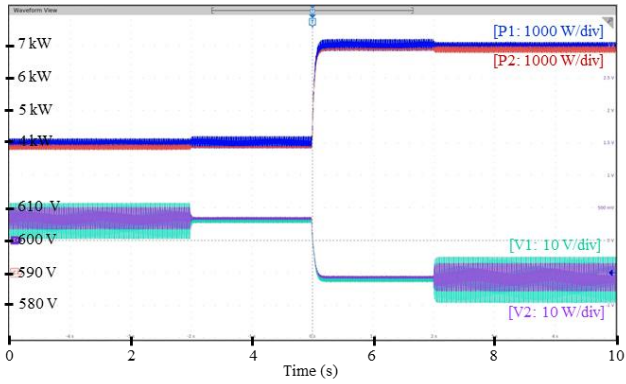

Fig. 10. HIL experimental results: power generation and output voltage of DERs

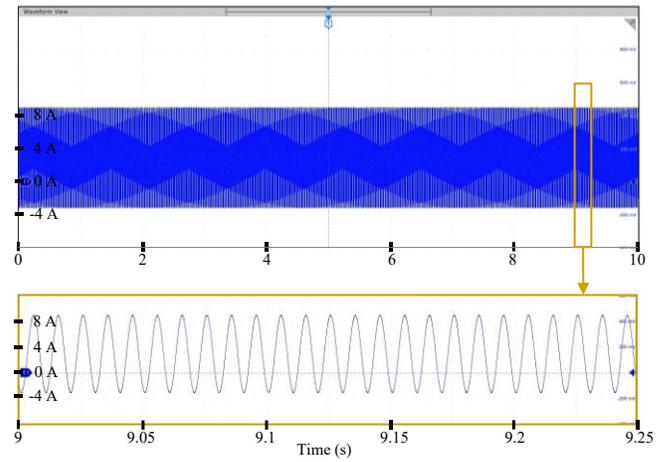

Fig. 12. HIL experimental results: DC side current of single-phase AC load ($i_{1\phi l,l}$).

The power generation of DERs is shown in Fig. 10. As illustrated, with the increase in the DC CPL at t=5s, the power generation of both DERs increases, indicating that the additional load is shared between them. Consistent with droop behavior, this increase in power is accompanied by a reduction in their terminal voltages (see Fig. 10). Figures 11, 12, and 13 show the output current of DER 1 ($i_1$), DC side current of the single-phase AC load ($i_{1\phi l,l}$), and the line 1 current ($i_{line,l}$), respectively. Before activation of the outer current loop (i.e., before $t = 3$ s), part of the harmonic current flows into DER 1, and the rest flows through line 1. The harmonic current is shared between them based on the harmonic impedance of DER 1 and the impedance seen at the connection point of Line 1. Consequently, during this period, the harmonic current injected into line 1 is uncontrolled. This harmonic current, propagated throughout the DC microgrid, causes a second harmonic voltage ripple at the terminals of both DERs (see Fig. 10). From t=3s to t=5s, once the outer loop is activated and DER1 plans to 100% compensation, the harmonic current of line 1 drops to zero (see Fig. 13), confirming the effectiveness of the proposed control method. It should be noted that for full harmonic compensation, $i_1^{ref,h}$ is set equal to $i_{l\phi l,l}$. By increasing the DC CPL at t=5s, still 100% compensation of the harmonic current is preserved, demonstrating the proper performance of the proposed control method even in the presence of disturbances. As illustrated in Fig. 10, eliminating the line harmonic current between t=3s and t=7s removes the voltage ripple. From t=7s to t=10s, $i_1^{ref,h}$ is set to $i_{l\phi l,l}/2$, resulting in 50% compensation of the load harmonic current. Figures 11-13 (more obvious in zoomed subfigures) show that half of the harmonic current now flows through line 1, while the other half is passed by DER 1. As shown in Fig. 10, the presence of harmonic current in the line during this partial-compensation interval leads to increased voltage ripple in both DERs, most noticeably at DER1.

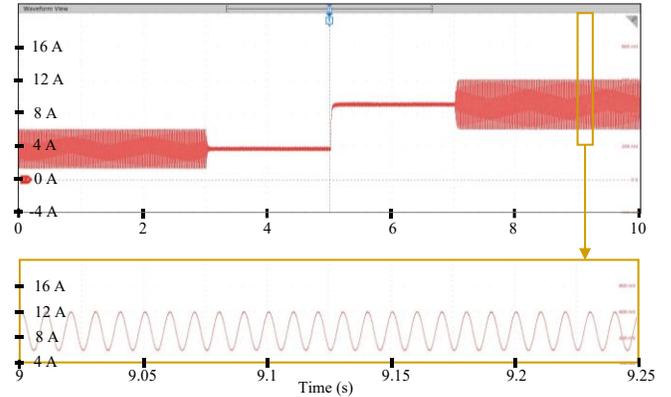

Fig. 13. HIL experimental results: line 1 current ($i_{line,l}$).

## VI. Conclusion

The instantaneous power of single-phase AC loads consists of a DC component (active power) and an oscillatory component, the frequency of which is twice the fundamental frequency of the AC voltage. In DC microgrids, single-phase AC loads are supplied through DC/AC converters. The oscillatory power of the AC load translates into a harmonic current with the same frequency in the DC side of the converter. When this harmonic current propagates through the DC microgrid, it increases the losses in the lines and gives rise to voltage ripples, which is characterized as a power quality issue and harms the loads. In this paper, this issue is resolved by ensuring that the load harmonic current is absorbed by the local DER. This way, the harmonic current does not propagate into the microgrid lines and hence the associated line losses and voltage harmonics are eliminated. To realize this objective, an HCM was proposed, in which the two control loops are utilized in parallel: 1) voltage control loop for regulating the DC component of the DER output voltage, 2) current control loop for controlling the harmonic component of the DER output current. This enables the DER to provide two functions at the same time, DC voltage regulation (in accordance with the droop characteristic) and harmonic current compensation (by setting the harmonic current reference equal to the single-phase load current). To prevent instability, the interaction between the two loops is minimized through utilization of different controllers for each loop. In particular, the voltage control loop employs a PI controller, whereas the harmonic current control loop employs a resonant controller. The PI controller has a high gain at DC but a negligible gain at the harmonic component. The resonant controller has a high gain at the harmonic frequency and zero gain at DC. This feature enables the implementation of the proposed control method without extracting the load harmonic current, which relaxes the computational burden.

The proposed control scheme is analyzed in frequency and time domains. The frequency domain analysis verifies that the current controller's closed-loop gain is 1 at the harmonic frequency (effectively tracking the harmonic current), and the voltage controller's closed-loop gain is 1 at DC (DC voltage regulation). Furthermore, the frequency domain results show that the voltage reference has a minimal influence on the output current at 100 Hz, and the current reference has a negligible influence on the voltage tracking at DC. This result demonstrates the interaction between the voltage and current control loops is minimal. HIL experiments have been conducted to validate the time domain response of the proposed control scheme. The HIL test results showcase the effectiveness of the proposed control method not only in terms of basic control functions (power sharing, voltage control) but also for harmonic current absorption and voltage quality improvement.

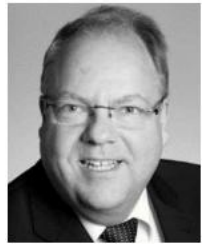

Thomas Ebel (Senior Member IEEE), received the Dipl.-Chem. degree (M.Sc. equivalent) in chemistry from Münster University, in 1992, and the Ph.D. degree (Dr.rer.nat.) from the Institute of Inorganic Chemistry, Münster University, in 1995. In 1995, he spent three months as a Guest Researcher with CNRS, Institute des Matériaux de Nantes, France, with Prof. J. Rouxel. From August 1995 to September 2001, he was a Research and Development Engineer and later the Research and Development Director of Siemens Matsushita Components, Siemens AG PR. Since October 1999, he has been with EPCOS AG, and since October 2008, he has been with TDK, Business Unit of Aluminum Electrolytic Capacitors, Heidenheim, Germany. From October 2001 to July 2008, he was the Research and Development Director, later the Technical Director (a CTO), and a member of the Board of Directors of Becromal Norway (Becromal S.p.A., since October 2008, Epcos, now TDK Foil), Milano, Italy. From September 2008 to July 2018, he was the Managing Director and a Shareholder with FTCAP GmbH (Husum Manufacturer of Aluminum Electrolytic and Film Capacitors), Germany. Since August 2018, he has been the Head of the Centre for Industrial Electronics (CIE), University of Southern Denmark (SDU), Sønderborg, Denmark. Since January 2022, he has been a Full Professor.

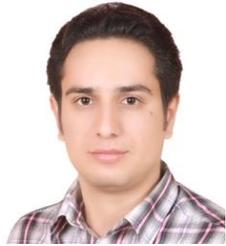

Mehdi Baharizadeh (Senior Member IEEE), received the B.Sc., M.Sc. (Hons.), and Ph.D degrees in electrical engineering from the Isfahan University of Technology, Isfahan, Iran, in 2008, 2011, and 2016, respectively. From 2017 to 2023, he was an assistant professor in the Department of Electrical Engineering, Khomeinishahr Branch, Islamic Azad University, Isfahan, Iran. Since 2023, he has been with Centre for Industrial Electronics, Institute of Mechanical and Electrical Engineering, University of Southern Denmark, where he is currently an assistant professor. His research interests include microgrids, energy transition, real-time digital Simulation, and power electronics.

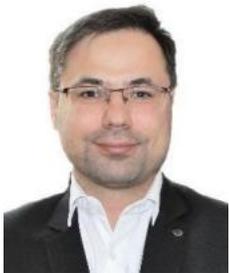

Mohammad Sadegh Golsorkhi (Senior Member IEEE), received the B.Sc. (Hons.) degree from Isfahan University of Technology, Isfahan, Iran, in 2009, the M.Sc. (Hons.) degree from Tehran Poly technique, Tehran, Iran, in 2012, and the PhD degree from The University of Sydney, Sydney, Australia, in 2016, all in electrical engineering. During 2015, he was a visiting PhD student with the Department of Energy Technology, Aalborg University, Denmark. In 2016, he worked in The University of Hong Kong, Hong Kong as a postdoctoral fellow. From 2017 to 2021, he was an Assistant Professor in the Department of Electrical and Computer Engineering, Isfahan University of Technology, Isfahan, Iran. Since 2021, he has been with the Centre for Industrial Electronics, Institute of Mechanical and Electrical Engineering, University of Southern Denmark, where he is currently an associate professor. His current research interests include microgrids, power electronics and grid integration of renewables.

Neda Keshavarzi received MSc in electrical engineering from Department of Electrical Engineering Khomeinishahr Branch, Islamic Azad University Isfahan, Iran in 2020. Here research interests are microgrids and control systems.